\documentclass[aip,cha,preprint,showpacs,showkeys,superscriptaddress,twocolumn,10pt]{revtex4-1}
\usepackage[T1]{fontenc}
\usepackage{graphicx}
\usepackage{color}
\begin{document}

%\preprint{}

\title{Organization and clasification of trajectories in  the time-delayed  Mackey-Glass system}



\author{Pablo Amil}
\affiliation{Facultad de Ciencias, Universidad
de la Rep\'ublica, Igua 4225, Montevideo, Uruguay}

\author{Cecilia Cabeza}
\affiliation{Facultad de Ciencias, Universidad
de la Rep\'ublica, Igua 4225, Montevideo, Uruguay}

\author{Cristina Masoller}
\affiliation{Departament de Fisica i Enginyeria Nuclear, Universitat
Politecnica de Catalunya, Colom 11, E-08222 Terrassa, Barcelona, Spain}

\author{Arturo C. Mart\'{\i}}
\affiliation{Facultad de Ciencias, Universidad
de la Rep\'ublica, Igua 4225, Montevideo, Uruguay}

%\thanks{}


\date{\today}

\begin{abstract}
Using a novel electronic implementation of a well-known time-delayed system, the Mackey-Glass (MG) system, we 
investigate the organization of the trajectories in the phase space, and classify the coexisting
solutions,  both, in observations and in model simulations. The numerical simultations are performed using  a
discrete-time equation that approximates the exact solutions of the MG model and in particular,
models the delay line in the electronic circuit. In wide parameter regions, different periodic or aperiodic solutions,
but with similar waveforms exhibiting the alternation of peaks of different amplitudes , coexist.
A symbolic algorithm is proposed to classify those solutions. 
The system's phase-space was explored by varying the parameter values of two  families  of initial functions. 
\end{abstract}


\maketitle

%\section{Introduction}

%\quotation{prueba}

\begin{quotation}
Multistability, i.e., the coexistence of several attractors for a given set of parameters,
is a characteristic feature of nonlinear systems, and in particular, of systems with time-delays.
One paradigmatic example of time-delayed system is the well-known  Mackey-Glass (MG) model which 
deals  with physiological processes, mainly respiratory and hematopoietic (\textit{i.e.} formation of blood cellular
components) diseases.  The MG's work had an impressive impact. 
Since its publication, it exhibits nearly 3000 cites in  scientific journals and, at present, Google reports more than two
millions results for the search \textit{Mackey-Glass}. 
In general, in time-delayed systems, the evolution of the
system at a given time not only depends on the state of the system at
the current time but also on the state of the system at
\textit{previous} times.  The dynamics of processes involving time delays red, as those studied by MG,
is far more complex than that of non-delayed, \textit{i.e.} instantaneous,
systems. Actually, if the dynamics of a  system  at time $t$ depends on the state of
the system at a previous time  $t-\tau$, 
the information needed to predict the evolution is contained in the entire interval $(t-\tau,t)$.  
Thus, the evolution of a delayed system depends on \textit{infinite} previous values of the variables.
In mathematical terms, delayed systems are modelled in
terms of delayed differential equations (DDEs) and one single DDE is equivalent to a infinite set of ordinaries differential 
equations. Here, using a novel electronic implementation of a well-known time-delayed system, the Mackey-Glass (MG) system, we 
investigate the organization of the trajectories in the phase space, and classify the coexisting
solutions,  both, in observations and in model simulations.

\end{quotation}

\section{\label{sec:intro}Introduction}

Chaotic systems are characterized by unpredictable behavior; however, it is well known that they exhibit
a certain degree of regularity and structure. Nonlinear systems often display multistability, that is, the coexistence
of different attractors for the same set of parameters. From observed noisy time-series that display similar oscillatory patterns,
identifying and distinguishing different coexisting attractors is a challenging task,
in particular when noise induces switching among different attractors. Recently, the phenomenon of extreme multistability
(i.e., the presence of an infinite number of attractors for a given set of parameters) has been predicted theoretically,
and observed experimentally \cite{hens2012obtain,patel2014experimental}. While in most multi-stable systems
chaotic attractors are rare \cite{feudel1997multistability}, systems with time delays are an exception to this rule.
A time-delay renders the phase space of a system infinite-dimensional, as one needs to specify, as initial condition,
the value of a function, $F_0$, over the time interval (-$\tau$,0), with $\tau$ being the delay time.
Time-delayed systems often display coexisting, high-dimensional attractors  
\cite{doyne1982chaotic,foss1996multistability,masoller1994coexistence,ahlers1998hyperchaotic,vicente2005analysis}.

Time-delays occur in a wide range of real-world systems, either due to couplings or to feedback loops (for recent reviews,
see \cite{just2010delayed,flunkert2013dynamics}), and many practical applications have been demonstrated, for example, it has been shown that delay-dynamical systems, 
even in their simplest manifestation, can perform efficient information processing \cite{appeltant2011information}, and the high dimensionality of the chaotic dynamics 
that they generate can be exploited for implementing ultra-fast random number generators \cite{uchida2008fast,kanter2010Optical}.

A paradigmatic time-delayed system is the Mackey-Glass (MG) model \cite{mackey1977oscillation},
which exhibits a rich variety of periodic and complex behaviors. 

The MG model, a first-order nonlinear delayed differential equation, was proposed in 1977 by Mackey and Glass to model physiological systems,
mainly respiratory and hematopoietic diseases (i.e. formation of blood cellular components)  \cite{mackey1977oscillation}.
The onset of these diseases is associated with alterations (bifurcations) in the periodicity of physiological  variables,
for example, irregular breathing patterns or fluctuations in peripheral blood cell counts \cite{de2000onset}. Specifically, Mackey and Glass considered
a population of mature circulating white blood cells and a delay, $\tau$, between the initiation of cellular production in the bone narrow and their release into the blood.
The dynamics of the density of blood cells is increasingly complex as the delay grows. When $\tau=0$ there is a stable equilibrium point 
which becomes unstable with increasing delay and periodic solutions appear. As $\tau$ increases a sequence of bifurcations generates oscillations with higher
periods and eventually aperiodic behavior.

Over the years the dynamics of the MG model has been investigated numerically 
\cite{ding2007stability,wei2007hopf,wan2009bifurcation,berezansky2012mackey,junges2012intricate,zhang2014almost}, 
and it has been used to generate high-dimensional chaotic signals \cite{grassberger1983measuring}. 
The MG model has also been used as a toy model to study chaos synchronization in delayed systems \cite{masoller2001anticipation,PhysRevE.74.016211,PhysRevE.75.016207}.
Several groups have proposed experimental implementations via electronic circuits 
\cite{namajunas1995electronic,kittel1998generalized,PhysRevE.75.016207,pham2012implementation,amil2014electronic}.
The scope of the present work is to take advantage of a recently proposed electronic implementation \cite{amil2014electronic} 
to investigate the coexistence of different, but very similar, high-dimensional periodic and aperiodic solutions.

First, we perform a critical comparison of model simulations and experimental observations.
We show that, in spite of the fact that in the electronic circuit the infinite phase space of the MG system
is discretized via a finite set of $N$ values of the initial function, $F_0$, if $N$ is large enough
the electronic circuit \cite{amil2014electronic} is indeed a highly precise implementation of the MG model.
Then, we study the multistability of coexisting solutions by distinguishing and classifying different periodic and aperiodic solutions, which display very
similar oscillatory patterns.%, and their frequency of occurrence, both, in observations and in model simulations.
%The excellent agreement found indicates that the electronic circuit is an accurate representation of the MG model, at least for the parameters considered here, and thus provides useful experimental tool for the analysis of other interesting features of the MG model, such as the interplay of noise, multistability and delay, or the variation of the number of coexisting attractors, and their frequencies of occurrence, with the delay.

This paper is organized as follows: in Sec.~\ref{sec:mg} the Mackey-Glass model and the electronic implementation are  described.
In Sec.~\ref{sec:method} the model equations are discretized and  in  Sec.\ref{sec:bifs} the experimental and numerical
bifurcation diagrams are compared in order to demonstrate that the electronic circuit perfectly reproduces the MG model. 
Then, in Section~\ref{sec:multi}, the coexistence of periodic and aperiodic solutions is analyzed.
Finally, Sec.~\ref{sec:con} presents a summary of the results and the conclusion.

\section{\label{sec:mg}Model and electronic circuit}

The Mackey-Glass delay-differential equation is \cite{mackey1977oscillation}
\begin{equation}
\frac{dP}{dt}=\frac{\beta_{0}\Theta^{n}P_{\tau}}{\Theta^{n}+P_{\tau}^{n}}-\gamma P
\label{eq:MG}
\end{equation}
where $P$ is the density of mature circulating white blood cells, $\tau$ is the delay time and $P_{\tau}=P(t-\tau)$. The parameters $\Theta$ and $\beta_{0}$ and the exponent $n$ are related to the production of white blood cells while $\gamma$ represents the decay rate.

The number of parameters can be reduced by re-scaling the variables $x= P/\theta$ and $t' = t  \gamma $. After 
re-scaling, the equation for $x\left(t'\right)$ reads as
\begin{equation}
\frac{dx}{dt'}=  \alpha\frac{x_{\Gamma}}{1+x_{\Gamma}^{n}}-x  % =   f(x_{\Gamma})-x
\label{eq:MGlinda}
\end{equation}
where $\Gamma = \gamma\tau$ is the normalized delay time, $\alpha = \beta_{0} / \gamma$,
%$f(x) = \alpha x /( 1+x^{n})$ 
and $x_{\Gamma}=x(t'-\Gamma)$.

\begin{figure}[h]
\begin{centering}
\includegraphics[width=0.99\columnwidth]{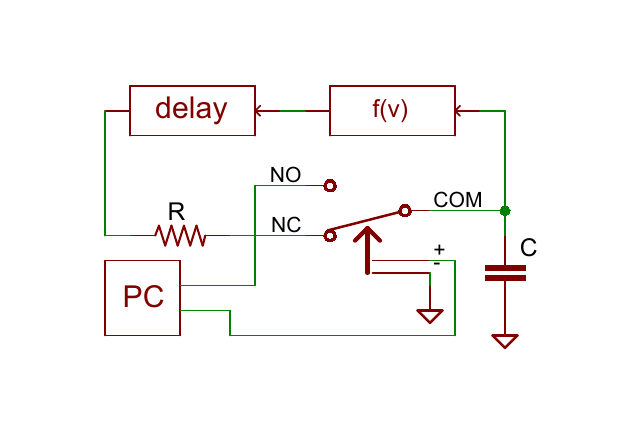}
\end{centering}
\caption{\label{fig:MGbloqRelay}
Experimental setup. The electronic circuit consists of two passive elements, $R$ and $C$, and two blocks, one implements the 
nonlinear function, $f(v)$, and the other the delay. The initial conditions  can be arbitrarily set by using  the relay  which allows to switch between
a free evolution in the normal closed (NC) position, and a controlled evolution in the normal open (NO) position.
The relay state and the voltage in the NC position are controlled by a PC.
}
\end{figure}

The electronic implementation of the MG model, as given by Eq.~(\ref{eq:MGlinda}), presents two main parts: the delay block, which produces only a time shift between the input and the output, and the function block, which implements
the nonlinear function. A schematic view of the circuit is shown in Fig.~\ref{fig:MGbloqRelay},
while a detailed description can be found in \cite{amil2014electronic}.

The purpose of the delay block is to copy its input as an output after some
delay time. The implementation of this block with analog electronic is possible by
using a Bucket Brigade Device (BBD), which is a discrete-time analog device.
Internally it contains an array of $N$ capacitors in which a signal travels one step at a time.
The origin of the name  comes from the analogy with the term \textit{bucket brigade},  used for a line of people passing buckets of water.
In this work, we used the integrated circuits MN3011 and MN3101 as BBD and clock signal generator respectively.

This approach for implementing a delay approximates the desired
transfer function given by $v_{out}\left(t\right)=v_{in}\left(t-\tau\right)$, by sampling the
input signal and outputting those samples $N$ clock periods later. Thus, if $dt$ is the clock period, the delay time is $\tau=Ndt$.
In the MN3011 $dt$ can vary between $5$~$\mu s$ and $50$~$\mu s$.

The number of capacitors, $N$, can be selected among the values provided by the manufacturer; for our devices, $N=396, 662, 1194, 1726, 2790,3328$.
The function block was implemented for an exponent $n=4$, however, other values of $n$ could be similarly implemented.
Integrated circuits AD633JN  and AD712JN  were used to implement sums, multiplications and divisions
because of their simplicity, accuracy, low noise and low offset voltage.

The initial conditions are given by the voltages in the capacitors, which are defined by an input signal of duration greater than the delay time.
A relay synchronized with an analog output of a PC was used as depicted in Fig.~\ref{fig:MGbloqRelay}. This setup allows to switch the system between a free evolution in the normal closed (NC) position of the relay, and a controlled evolution in the normal open (NO) position.
In this way, with the relay in the NO position, the initial values stored in the delay line (defined by the input signal) were fully controlled by the PC. Then, the relay was set to the NC position, the circuit evolved freely, we measured the voltage immediately after the delay block,
and reconstructed the voltage in the capacitor with a tuned digital filter.

A remarkable advantage of this electronic implementation is that different input signals (i.e., different functions of time)
may be selected as initial conditions. As the initial function is fully controlled by the computer,
it is easy and straightforward to analyze the evolution of the MG system starting from arbitrary initial conditions.

By applying Kirchhoff's laws to the circuit shown in Fig.~\ref{fig:MGbloqRelay} with the relay in the NC position,
the equation describing the voltage at the capacitor terminals, $v$, is
\begin{equation}
\frac{dv}{dt}=\frac{1}{RC}\left[f(v_{\tau})-v\right]
\label{eq:ElecEq}
\end{equation}
where $v_{\tau}=v(t-\tau)$ and $f\left(v\right)=\beta\frac{v}{\theta^{n}+v^{n}}$, with $\beta$ and $\theta$ circuit constants. 
Using the definition of the dimensionless time  $t'= t \gamma$, setting  the characteristic time-scale of the system as $RC=\gamma^{-1}$,
and $f$  as the nonlinear function of the MG model,
this equation can be identified with Eq.~(\ref{eq:MGlinda}).

%\section{Numerical methods and comparison}
\section{Model discretization}
 \label{sec:method}

%\textcolor{red}{In this Section we present the numerical methods and compare the numerical with the  experimental results. The  numerical results are based on a discretization of the equation governing the dynamics of the MG circuit (see \cite{amil2014electronic} for additional details). }

To analyze whether the electronic circuit described in the previous section indeed represents the MG model
(i.e., to assess the impact of the implementation of the delay via an array of $N$ capacitors in the Bucket Brigade Device)
we first discretize the MG delay-differential equation (as in Eq.~\ref{eq:ElecEq}) and then compare the simulations of 
the discretized MG model with observations from the electronic circuit.

%To obtain the discrete equation we note that, since the BBD is a discrete-time analog device,
%the output of the delay block remains constant during  each clock period. Then, the delayed value, $v_{c\tau}$, can be approximated
%by the value at the beginning of the sampling period
%\begin{equation}
%v_{c,\tau}  \simeq v_{c}\left(dt\left\lfloor \frac{t}{dt}-N+1\right\rfloor \right)
%\label{eq:circuit1},
%\end{equation}
%where the symbols $\lfloor .\rfloor$ stand for the integer part function.
The usual way to discretize a delay-differential equation is to approximate the delayed term $f( v_{\tau} )$ as constant in 
the small time interval $(t, t + dt)$. In this way, one can integrate Eq.~(\ref{eq:ElecEq}) and obtain
\begin{equation}
v(t+dt) =  [ v(t) - f( v_{\tau} ) ] \mathrm{e}^{-dt /RC } + f( v_{\tau} ).
\label{eq:dt}
\end{equation}
Denoting $t=jdt$, $\tau=Ndt$, $v_{j}=v(t)$, and $v_{j-N+1}=v_{\tau}=v(t-\tau)$, Eq.~\ref{eq:dt} reads
\begin{equation}
%v_{j+1} =  \left(v_{j}  -  f\left(v_{j-N+1}\right)\right)    \mathrm{e}^{-dt /RC }+f\left(v_{j-N+1}\right).
v_{j+1} =  \left[v_{j}  -  f\left(v_{j-N+1}\right)\right]   \mathrm{e}^{-dt /RC }+f\left(v_{j-N+1}\right).
\label{eq:iterlinda}
\end{equation}
%Substituting the expression for the nonlinear function $f(v)$  it results
%\begin{equation}
%v_{j+1}=   v_{j} \mathrm{e}^{-dt /RC }     + (1 -  \mathrm{e}^{-dt /RC } )   \frac{   \alpha v_{j-N+1}}{ 1 +v_{j-N+1}^{n}}.
%\label{eq:iterlinda}
%\end{equation}
The solutions of this discrete-time equation tend to the solutions of the original delay-differential equation, Eq.~(\ref{eq:ElecEq}), 
as $dt$ tends to zero and $N$ grows to infinity while the delay time $\tau=N  dt$ is kept constant. To test the agreement with the electronic implementation, 
in the next section we compare the solutions of Eq. (\ref{eq:iterlinda}) with observed time-traces from the electronic circuit, using the smallest possible $N$ value
(396) in order to consider the worst-case situation, because, if a good agreement is found with this value, then, an even better agreement can be expected for larger $N$. 
Clearly, here the issue is to analyze how good the approximation of the delayed term $f( v_{\tau} )$ as a constant value in the interval $(t, t + dt)$ is, given the 
specific characteristic time-scales of the electronic circuit.
%Eq. \ref{eq:ElecEq} is simulated using a this value.

\section{Results}

\subsection{Experiment-model comparison}
\label{sec:bifs}

\begin{figure}[h]
\begin{centering}
\includegraphics[width=0.99\columnwidth]{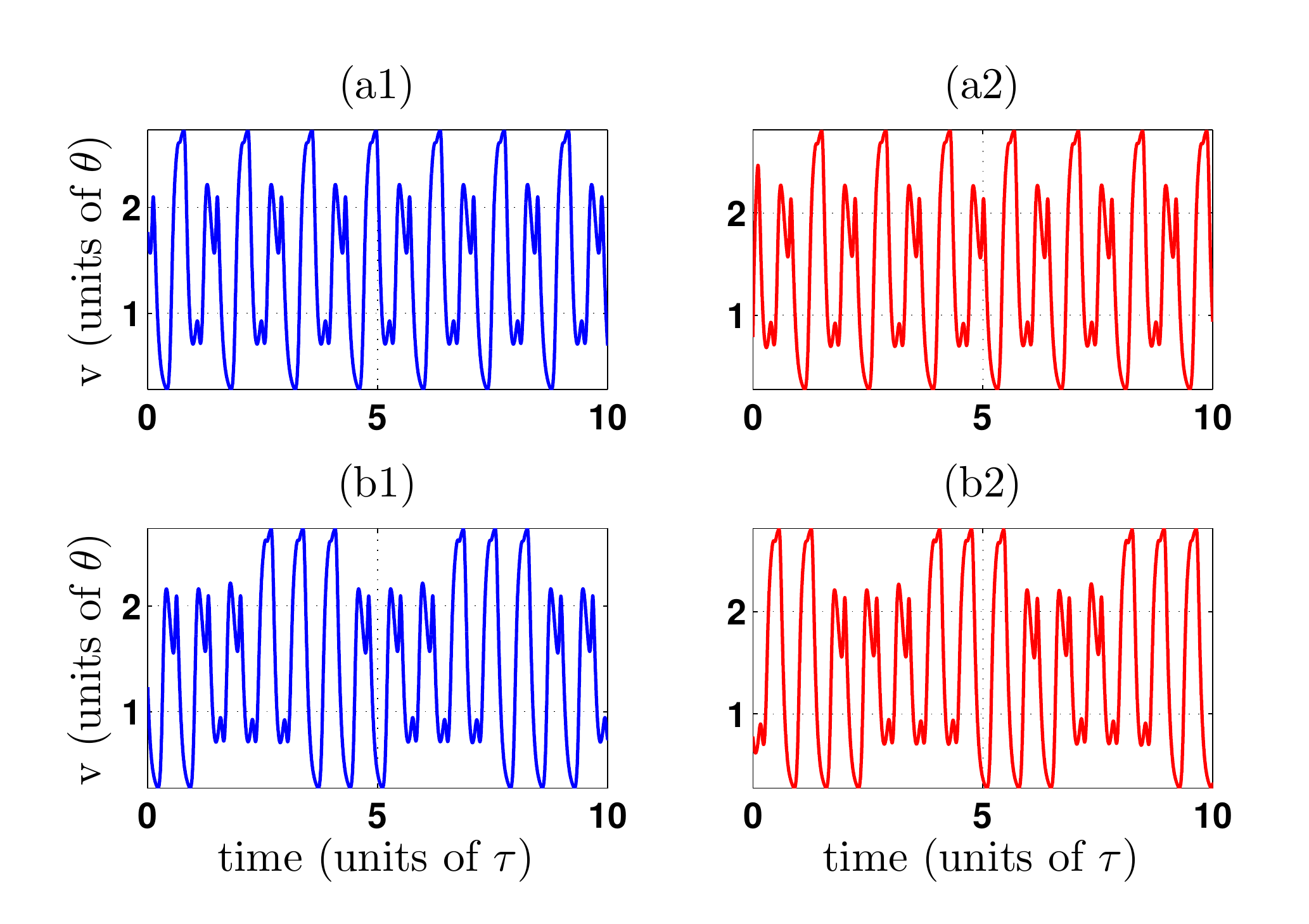}
\end{centering}
\caption{\label{fig:Sum3Series} Comparison between simulated (left column) and experimental (right column) time series.
The top and bottom rows display coexisting solutions obtained from different initial functions. The parameter values are:
$n=4$, $\alpha=4.9$, $\Gamma=\tau/RC=15.7$ and $N=396$. %For these parameters there is an excellent agreement experiments-simulations. %Note the alternations of maxima of different amplitudes in each solution.
}
\end{figure}

Figure~\ref{fig:Sum3Series} displays two examples of temporal evolutions, one is synthetic, obtained from simulations of Eq.(\ref{eq:iterlinda}), 
and the other is empirical, recorded from the electronic circuit (the initial conditions are as described in Sec.~\ref{sec:multi}). One can 
notice that there is an excellent agreement experiments-simulations: two coexisting solutions were found, both, in the simulations and in 
the electronic circuit, which are characterized by the same alternation of peaks of different amplitudes.
%\textcolor{red}{The excellent agreement between the numerical and experimental resuts is revealed. The time-shifted cross-correlation (not shown here) also corroborated the quantitative correspondence between experiments and simulations}.

\begin{figure}
\begin{centering}
\includegraphics[width=0.99\columnwidth]{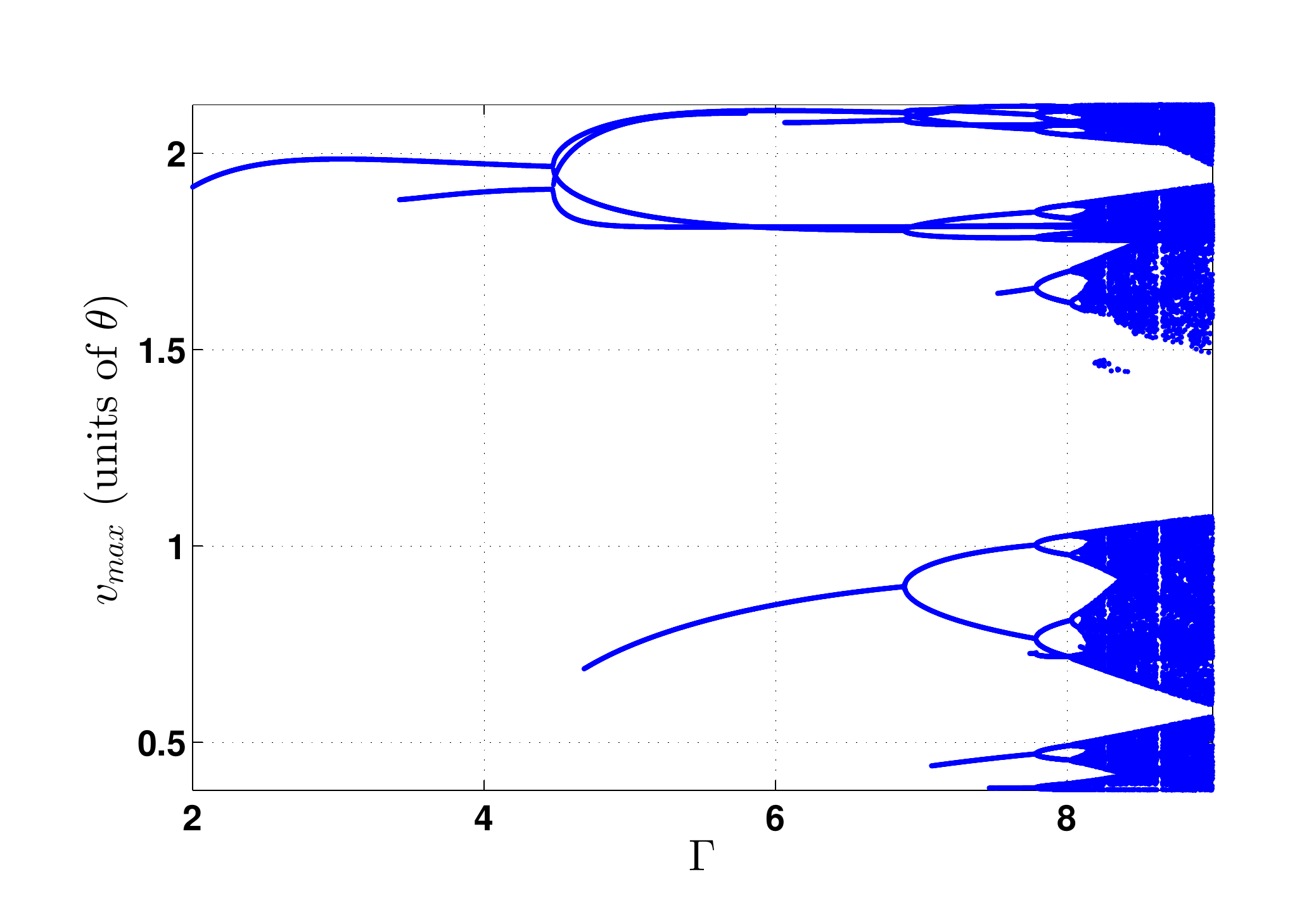}
\end{centering}
\caption{
\label{fig:D-TBif}
Bifurcation diagram displaying the maxima of time series, as a function of the normalized delay, $\Gamma=\tau/RC=Ndt/RC$, 
obtained from simulations of Eq.~(\ref{fig:D-TBif}). $\alpha=3.73$, other parameters are as in Fig. ~\ref{fig:Sum3Series}.}
\end{figure}

\begin{figure}
\begin{centering}
\includegraphics[width=0.99\columnwidth]{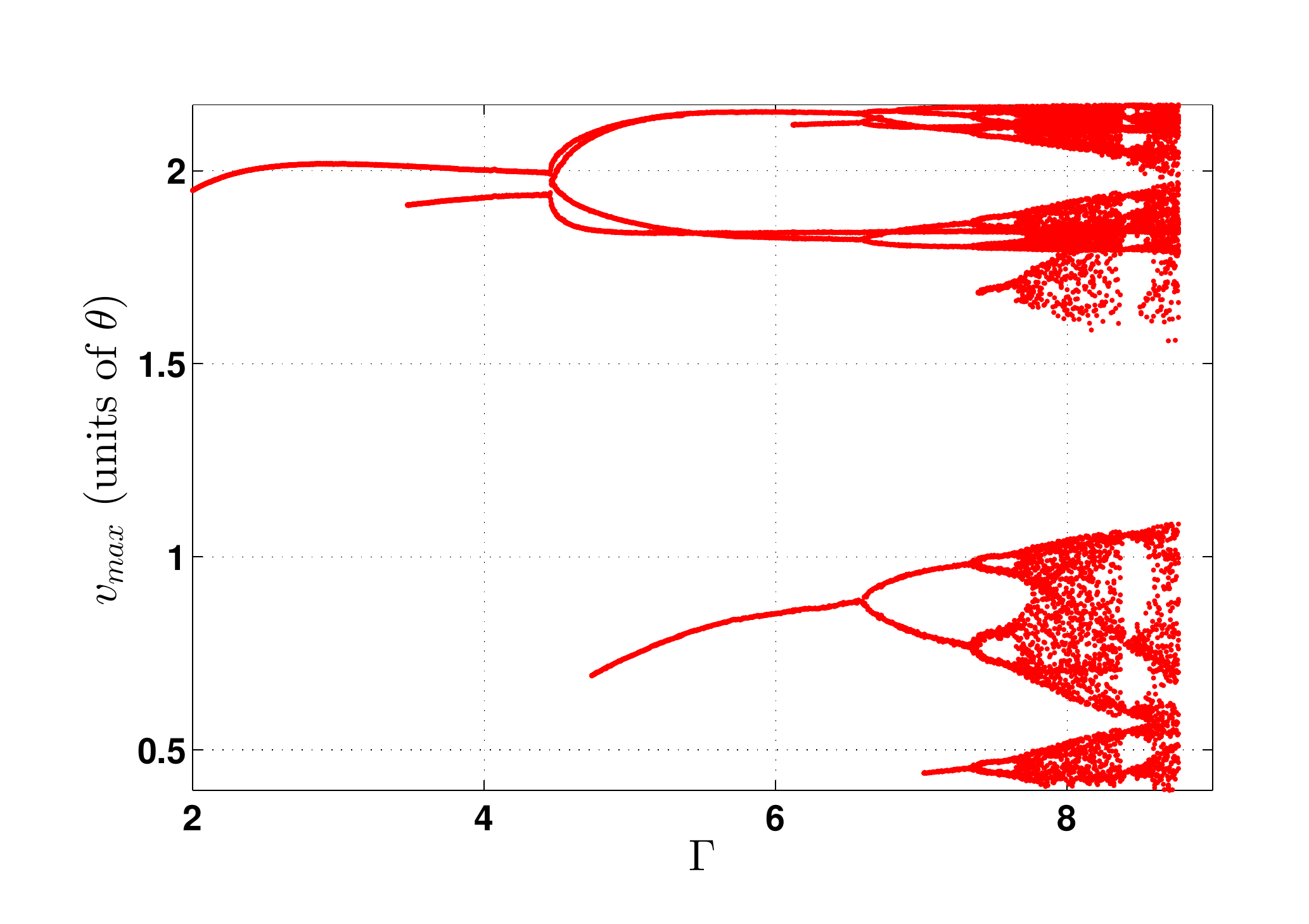}
\end{centering}
\caption{ \label{fig:ExpBif}
Empirical bifurcation diagram. The electronic circuit delay line has $N=1194$ $C=1.0$~$\mu$F capacitors. To vary
$\Gamma=\tau/RC=Ndt/RC$ within the same range as Fig.~\ref{fig:D-TBif}, $R$ was varied in the range 0.5 k$\Omega$ - 1.0 k$\Omega$ and $\alpha=3.73$, as in Fig.~\ref{fig:D-TBif}.}
\end{figure}

Figures \ref{fig:D-TBif} and \ref{fig:ExpBif} display bifurcation diagrams and demonstrate that the agreement is very good for
a wide range of normalized delays. These bifurcation diagrams were obtained by plotting, after neglecting transients, the 
maxima of time series, as a function of $\Gamma=\tau/RC$. In the experimental setup, $\Gamma$ was varied by changing $R$ (i.e., $C$, $N$ and $dt$ were kept fixed).
 %On the other set of measurements, to be discussed in the next subsection, successive temporal evolutions were obtained without varying any parameter, but, varying  the initial conditions of the system according to a prescribed rule.

%We can again observe an excellent agreement of the simulations, Fig. \ref{fig:D-TBif}, and the experiments, Fig. \ref{fig:ExpBif}.
In addition to the familiar period-doubling and chaotic branches, singles branches that appear or disappear at certain $\Gamma$ values can be appreciated. These isolated  branches are  typical  of  delay-differential equations \cite{junges2012intricate}.  We observe here that they appear
at the same value of $\Gamma$, both, in the  experimental and in the numerical diagram. We can also note that, for the highest  $\Gamma$ values, the numerical 
and experimental diagrams show a few small differences; this can be expected because the approximation used to derive  Eq.~(\ref{eq:iterlinda}) [\emph{i.e.}, $f(v_{\tau})$ is constant in $(t, t + dt)$], worsens.

\subsection{Analysis of multistability}
\label{sec:multi}

\begin{figure*}[htb]
\begin{minipage}[b]{0.49\linewidth}
\centering
\includegraphics[width=0.98\columnwidth]{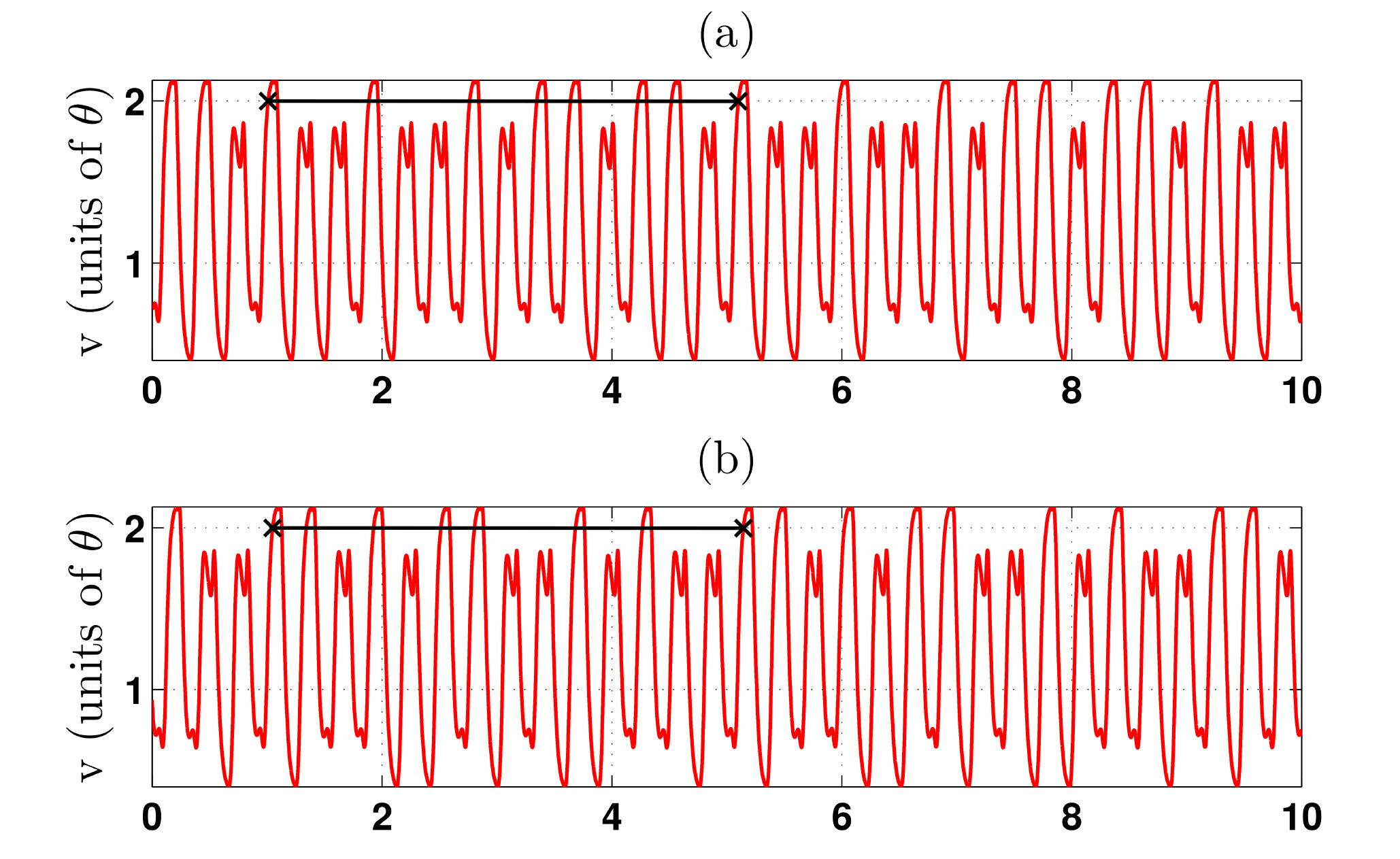}
\includegraphics[width=0.98\columnwidth]{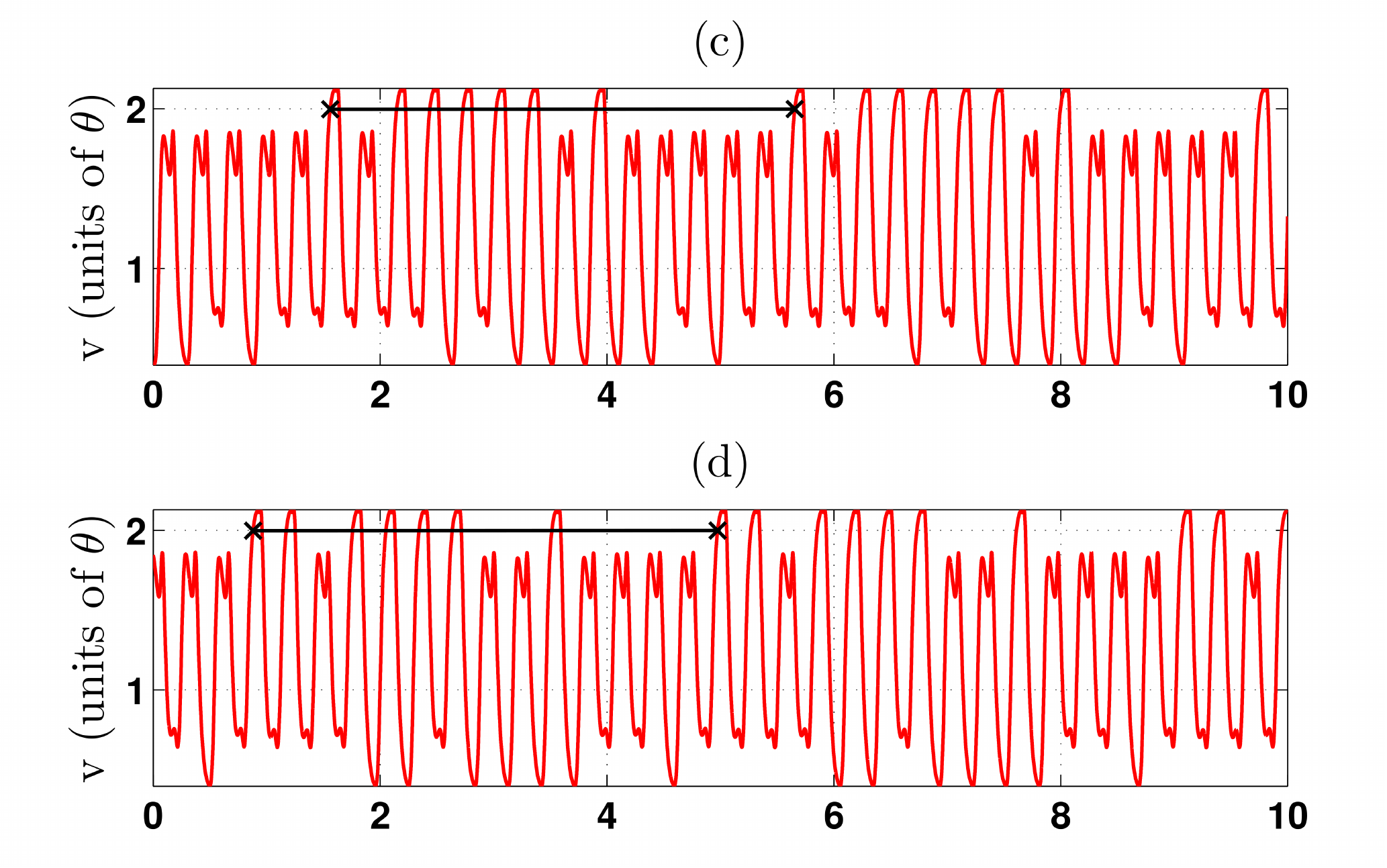}
\end{minipage}
\begin{minipage}[b]{0.49\linewidth}
\centering
\includegraphics[width=0.98\columnwidth]{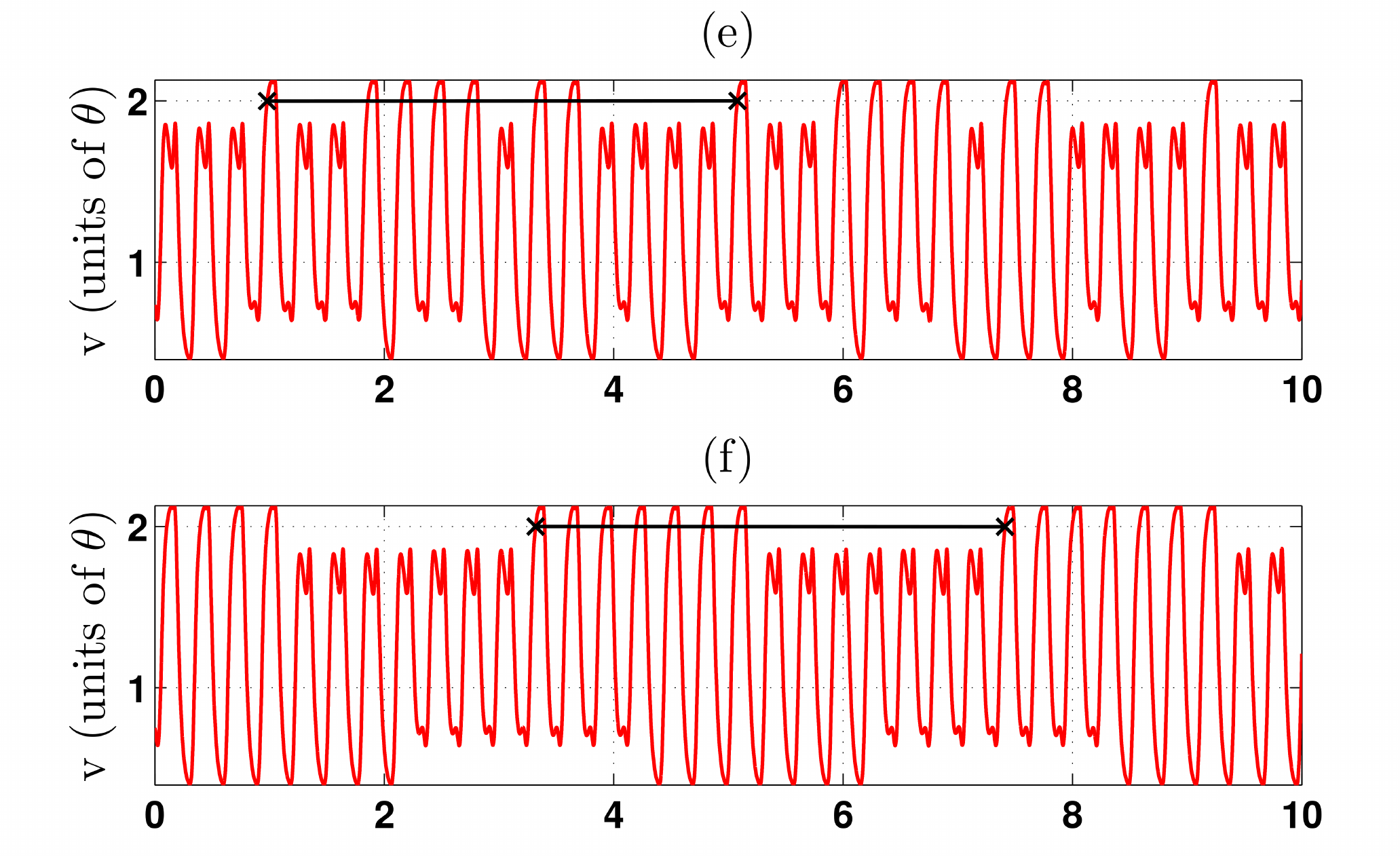}
\includegraphics[width=0.98\columnwidth]{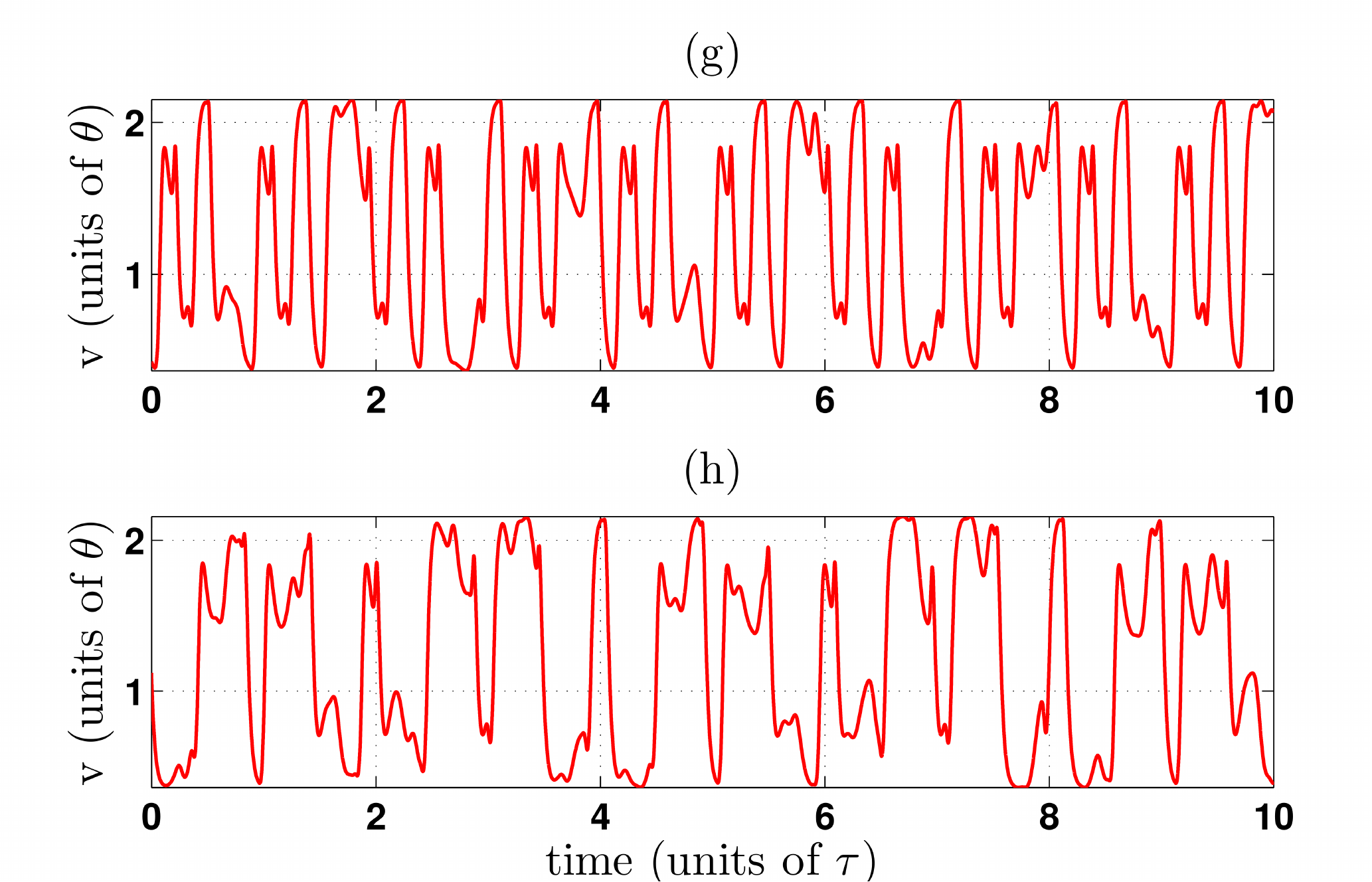}
\end{minipage}%
\caption{\textcolor{red}{Periodic and aperiodic experimental time series. The period, if exists, is indicated with a black line.}
The parameters of the electronic circuit are $n=4$, $\alpha=3.71$, $\Gamma=40$. The initial condition is in Eq.~(\ref{Fphi}).
\label{fig:temporalseries}
}
\end{figure*}

Next we investigate the influence of the initial conditions. As it is well known, time-delayed systems often display similar coexisting solutions.
In order to identify parameter regions where multi-stability occurs, we developed an algorithm for time-series analysis that allows to unambiguously 
distinguish similar waveforms.
%In general, when the control parameter and the initial conditions  are varied, similar solutions, periodic or aperiodic, with  different periods or number of peaks appear.
In Fig.~\ref{fig:temporalseries} several examples of empirical time-traces (recorded keeping constant the parameters of the electronic circuit) 
are shown: %experimental solutions for the same parameter values and varying the initial conditions are shown.
%In this case, 
six are periodic, (a)-(f), and two are aperiodic, 
%On the other hand, when a period was not detected the solutions are labeled as chaotic or aperiodic.
%To ilustrate this situation, six examples are shown in the panels  labeled 
(g) and (h). 
%Notice that, as the model does not incorporate explicitly external noise, the hypothesis of noise-induced switching among coexisting periodic solutions \cite{masoller2007polarization}  cannot be excluded.}
%For all the periodic temporal series displayed in Fig.~\ref{fig:temporalseries}, 
In the periodic time traces, the period (indicated with a black line), is $4.1 \tau$ and each period contains precisely 
$35$ maxima. It is remarkable that the number of maxima per period does not uniquely determine the solution (i.e., different coexisting solutions have 35 maxima per period).

The analysis algorithm is based in a symbolic representation of a time-series, and allows to label the different \emph{periodic} solutions.
%Such algorithm was robust enough to be used in both, experimental and numerical data.
%It, roughly, consisted on a symbolic analysis of the data, 
Two symbols were used, which correspond to highest peaks, and to 2nd highest peaks. %(which can be seen to be enough for the series we analyzed). 
Once the symbolic string was generated, the algorithm searched for periodicity, and if found, the time-series was labeled with the symbolic string, written in a unique way under cyclic permutations.
For example, the symbolic strings $AAABBBAAABBB$ and $BBBAAABBBAAA$ both represent the same periodic solution, which has three consecutive high maxima followed by three consecutive smaller maxima. %and as chaotic when periodicity was not found.
This way of labeling different solutions allows to distinguish among solutions with the same number of peaks per period. The algorithm can be extended to analyze more complex waveforms.

To investigate multi-stability one needs to consider different initial functions, $v(t-\tau)=F_0(t)$ with $t\in (-\tau,0)$. 
Here we consider two families with two parameters each: 
\begin{equation}
F_0(t)=\left(v_{2}-v_{1}\right) \frac{t}{\tau} +v_{2},
\label{Fv1v2}
\end{equation}
and
\begin{equation}
 F_{0}( t)= \frac{1}{40}\sin\left( \frac {7 \pi t}{2 \tau}  +\phi\right)\sin\left(  \frac {7 \pi t}{ \tau}  +2\phi\right)+v_{\mathrm{off}}.
 \label{Fphi}
\end{equation}
where ($v_{1}$, $v_{2}$) and ($\phi$,$v_{\mathrm{off}}$) univocally determine $F_0$ in the interval $ (-\tau, 0)$.

Then, for each pair of values, ($v_{1},v_{2}$) or ($\phi$,$v_{\mathrm{off}}$), a transient time is neglected (about $5000 \tau$ in the simulations and $1000 \tau$ in the experiments) and time series of length $200\tau$ (simulations) or $100\tau$ (experiments) are recorded. Their periodicity is analyzed with the symbolic algorithm and the solutions are plotted in the ($v_{1},v_{2}$) or ($\phi$,$v_{\mathrm{off}}$) plane.
If the MG system is only two-dimensional these plots would identify the basins of attraction of the different solutions; however, the MG system is a delayed system and thus, these plots only classify the different solutions obtained in terms of the two parameters that determine the initial function.

\begin{figure}[ht]
\begin{centering}
\includegraphics[width=0.98\columnwidth]{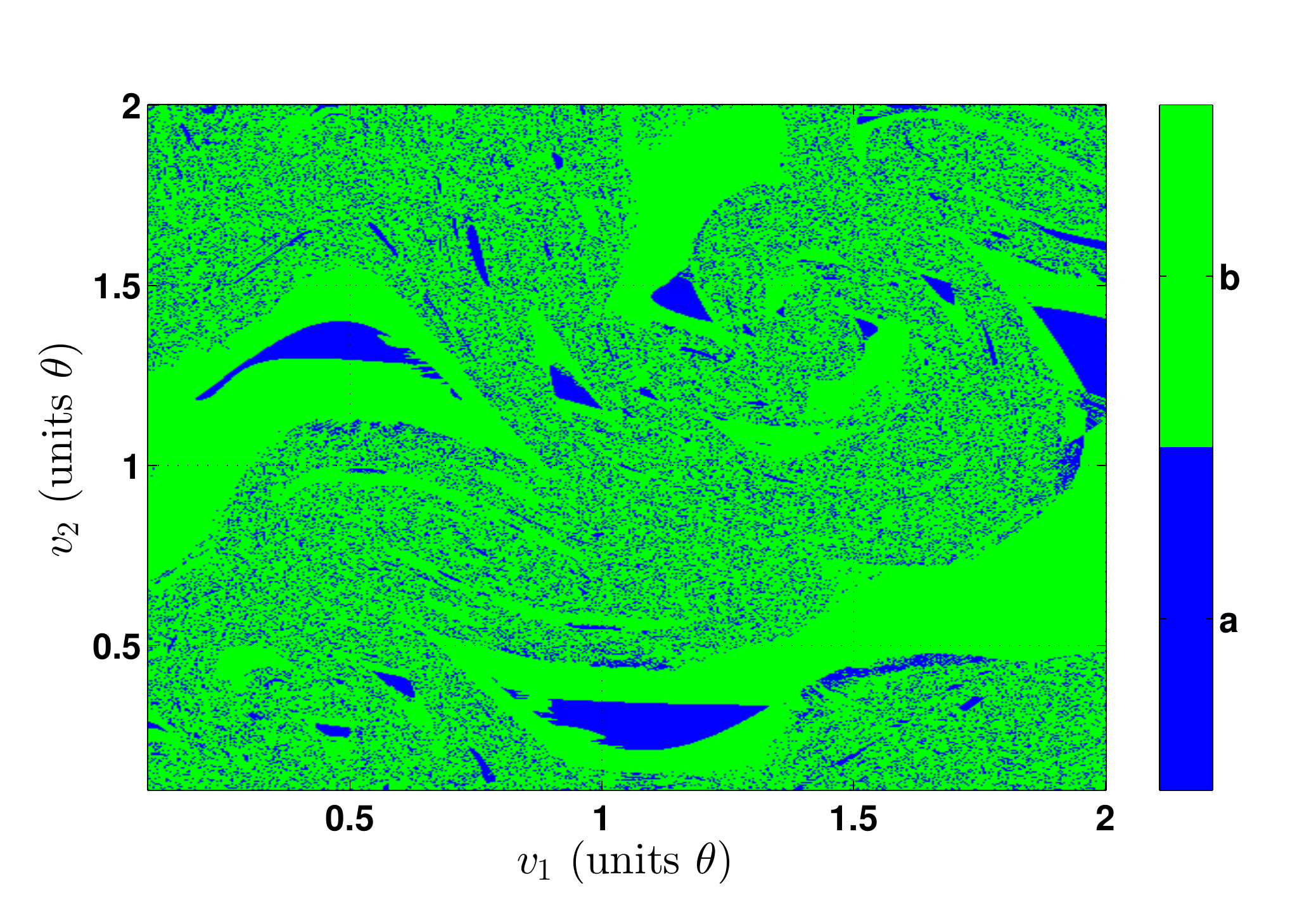}
\includegraphics[width=0.98\columnwidth]{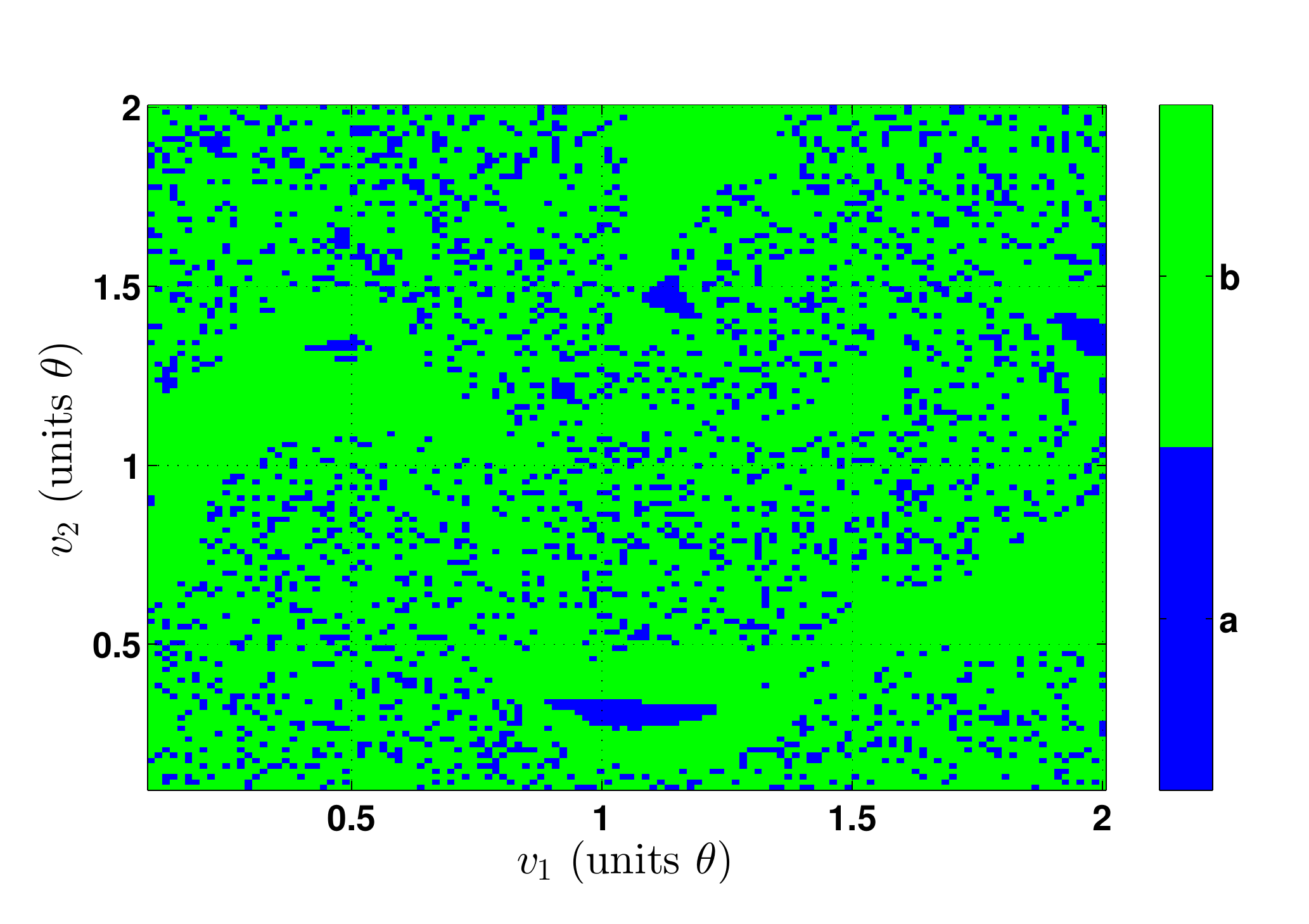}
\end{centering}
\caption{\label{fig:CondInic1} (Color online)
{
Map of parameters ($v_{1}$, $v_{2}$) [that define the initial function given by Eq.~(\ref{Fv1v2})], which evolve into one of two possible periodic solutions, \emph{a} in light grey (blue online) or \emph{b} in dark gray (green online). The corresponding waveforms are displayed in Fig.~\ref{fig:Sum3Series}.
The top panel displays the analysis of simulated time-series while the bottom panel, of empirical data. Parameters are as in Fig.~\ref{fig:Sum3Series}.}
}
\end{figure}

\begin{figure}[th]
\begin{centering}
\includegraphics[width=0.99\columnwidth]{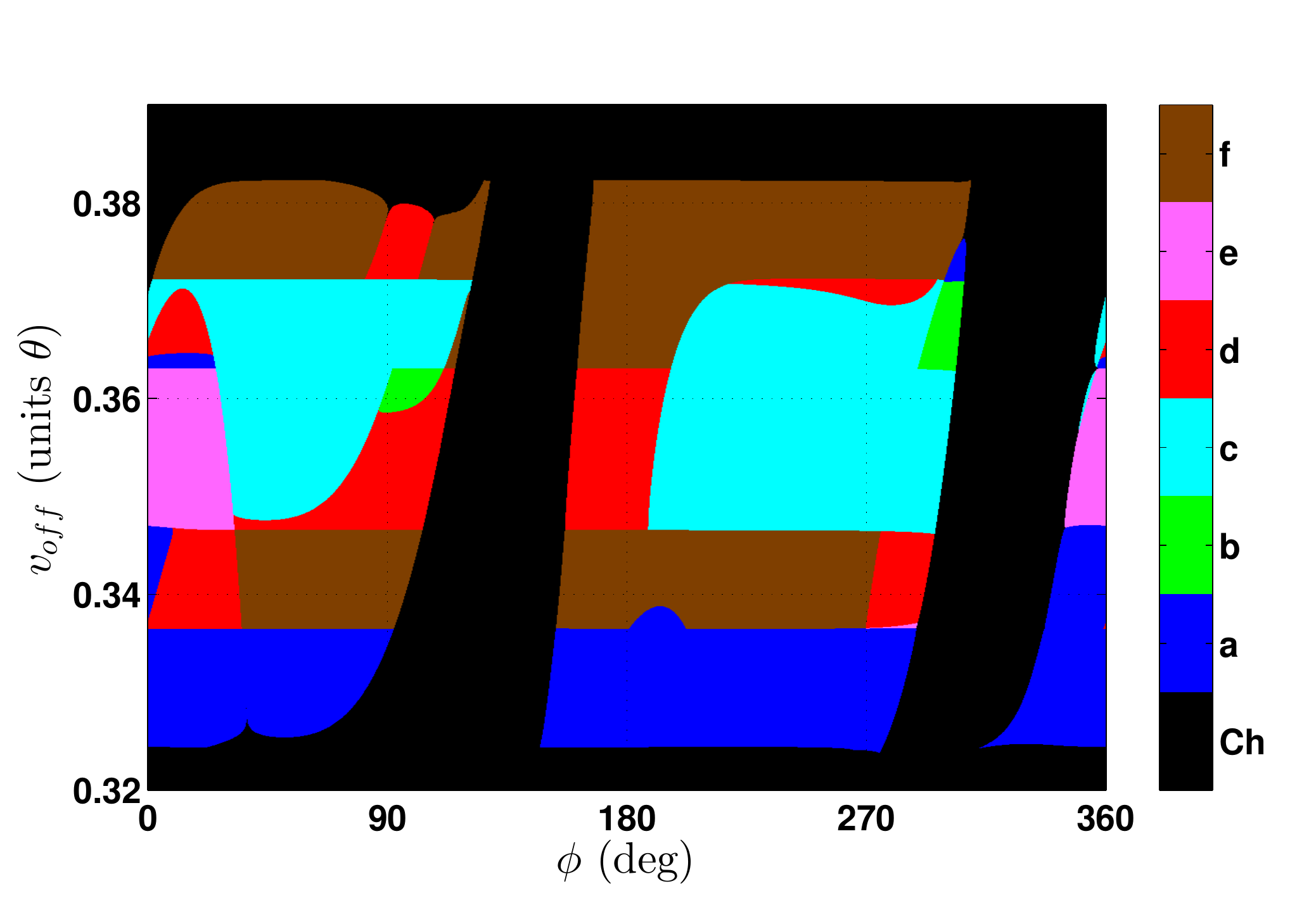}
\includegraphics[width=0.99\columnwidth]{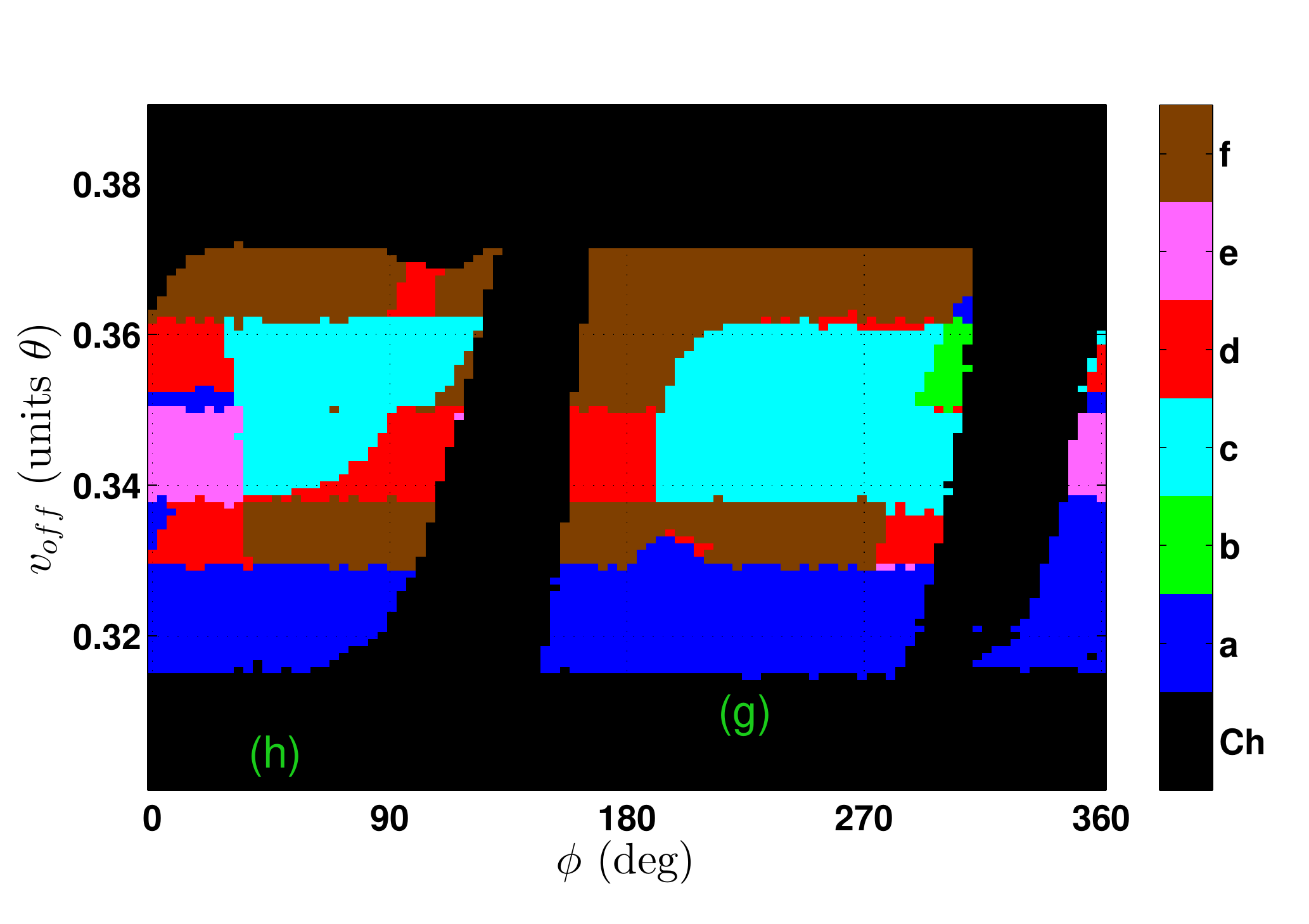}
\end{centering}
\caption{(Color online)
\label{fig:CondInic2}
As Fig. \ref{fig:CondInic1}, but now the initial function is defined by  Eq.~\ref{Fphi}, and the parameters are as in Fig.~\ref{fig:temporalseries}. 
The top panel displays the analysis of simulated time-series while the bottom panel, of empirical data. In both cases six different periodic waveforms were
identified, which are indicated in the color bar with letters \textit{a, b, c, d, e, f} corresponding to the panels in Fig.~\ref{fig:temporalseries}. Aperiodic
behavior is indicated in black. The labels \textit{g} and  \textit{h} in the map locate the initial conditions that generate trajectories as those displayed in  
Fig.~\ref{fig:temporalseries}, panels (g) and (f).
}
\end{figure}

The results are presented in Fig.~\ref{fig:CondInic1} (where $F_0$ is given by Eq.~\ref{Fv1v2} and the parameters of the MG model and of the electronic circuit 
are as in Fig. ~\ref{fig:Sum3Series}) and in Fig.~\ref{fig:CondInic2} (where $F_0$ is given by Eq.~\ref{Fphi} and the parameters are as 
in Fig.~\ref{fig:temporalseries}). In the first case there is bistability while in the second case, six different periodic 
solutions were identified (in Fig.~\ref{fig:CondInic2} the black regions represent initial functions that result in aperiodic trajectories). 
Experiments and simulations are contrasted, and again a very good agreement is found. Moreover, we computed the frequency of 
occurrence of the different coexisting solutions and again a very good agreement was found (not shown). Therefore, our study 
indicates that, at least for the model parameters considered here, the electronic circuit reproduces the main features of the MG 
system (the shape of the waveforms, the bifurcation diagrams, and the maps of bistable and multistable solutions) and thus, it 
could be used to investigate other issues, for example, noise-induced switching, or how multi-stability affects synchronization.

\section{Conclusion}
\label{sec:con}

Multistability in the Mackey-Glass (MG) model was studied experimentally, by using a novel electronic implementation, and numerically, by using a discrete-time equation that approximates the exact solutions of the MG model and in particular, models the delay line in the electronic circuit, which is implemented via a linear array of capacitors (a Bucket Brigade Device, BBD). We have found an excellent agreement between observations of the electronic circuit and the simulations of the discrete-time MG model.

In wide parameter regions, different periodic or aperiodic solutions, but with similar waveforms, coexist. In this work, these solutions, exhibiting the alternation of peaks of different amplitudes, were distinguished by means of a symbolic algorithm.

The system's phase-space was explored by varying the parameter values of two  families  of initial functions. The maps of initial conditions that result in different periodic solutions were found to exhibit complex structures, which are not uncommon in delayed systems \cite{shrimali2008nature}.
A full characterization of the complex organization of these solutions in the system's phase space is an open issue which deserves further research.

The electronic circuit investigated here can be a useful experimental tool for further understanding the bifurcation scenario and the complex solutions of the MG model, and can also be used as ``toy model'', to study generic features of time-delayed systems, such as deterministic high-dimensional attractors, synchronization in the presence of multistability, or the complex stochastic dynamics that can emerge due to the interplay of multistability, noise, and delay.

\begin{acknowledgments}

We acknowledge financial support from Programa de Desarrollo de las Ciencias Básicas (PEDECIBA),
Uruguay.  C.M. acknowledges financial support of grant FIS2012-37655-C02-01 of the Spanish Ministerio de Ciencia e Innovaci\'on.
\end{acknowledgments}

\bibliographystyle{aipnum4-1}
\bibliography{mybib}

\end{document}